\documentclass[conference]{IEEEtran}
\IEEEoverridecommandlockouts
\usepackage{cite}
\usepackage{amsmath,amssymb,amsfonts}
\usepackage{algorithmic}
\usepackage{graphicx}
\usepackage{textcomp}
\usepackage{xcolor}
\usepackage{enumitem}
\usepackage{array}
\usepackage{url}
\newcolumntype{P}[1]{>{\centering\arraybackslash}p{#1}}
\def\BibTeX{{\rm B\kern-.05em{\sc i\kern-.025em b}\kern-.08em
    T\kern-.1667em\lower.7ex\hbox{E}\kern-.125emX}}
\begin{document}

\renewcommand\IEEEkeywordsname{Keywords}

\title{%
  Enhancing GraphQL Security by Detecting Malicious Queries Using Large Language Models, Sentence Transformers, and Convolutional Neural Networks}

\author{
\IEEEauthorblockN{Irash Perera}
\IEEEauthorblockA{
\textit{Department. of CSE} \\
\textit{University of Moratuwa} \\
Colombo, Sri Lanka \\
irash.21@cse.mrt.ac.lk
}
\and
\IEEEauthorblockN{Hiranya Abeyrathne}
\IEEEauthorblockA{
\textit{Technical Lead Engineering} \\
\textit{WSO2} \\
Colombo, Sri Lanka \\
hiranyaa@wso2.com
}
\and
\IEEEauthorblockN{Sanjeewa Malalgoda}
\IEEEauthorblockA{
\textit{Director Engineering} \\
\textit{WSO2} \\
Colombo, Sri Lanka \\
sanjeewa@wso2.com
}
\and
\IEEEauthorblockN{Arshardh Ifthikar}
\IEEEauthorblockA{
\textit{Technical Lead Engineering} \\
\textit{WSO2} \\
Colombo, Sri Lanka \\
arshardh@wso2.com
}
}

\maketitle

\begin{abstract}
GraphQL's flexibility, while beneficial for efficient data fetching, introduces unique security vulnerabilities that traditional API security mechanisms often fail to address. Malicious GraphQL queries can exploit the language's dynamic nature, leading to denial-of-service attacks, data exfiltration through injection, and other exploits. Existing solutions, such as static analysis, rate limiting, and general-purpose Web Application Firewalls, offer limited protection against sophisticated, context-aware attacks. This paper presents a novel, AI-driven approach for real-time detection of malicious GraphQL queries. Our method combines static analysis with machine learning techniques, including Large Language Models (LLMs) for dynamic schema-based configuration, Sentence Transformers (SBERT and Doc2Vec) for contextual embedding of query payloads, and Convolutional Neural Networks (CNNs), Random Forests, and Multilayer Perceptrons for classification. We detail the system architecture, implementation strategies optimized for production environments (including ONNX Runtime optimization and parallel processing), and evaluate the performance of our detection models and the overall system under load. Results demonstrate high accuracy in detecting various threats, including SQL injection, OS command injection, and XSS exploits, alongside effective mitigation of DoS and SSRF attempts. This research contributes a robust and adaptable solution for enhancing GraphQL API security.
\end{abstract}


\section{Introduction and Background}
The adoption of GraphQL has grown due to its efficiency in allowing clients to request specific data, which optimizes data transfer. However, this flexibility introduces unique security challenges, as the dynamic nature of GraphQL queries makes them susceptible to attack vectors that conventional security measures for static APIs often miss.

Malicious actors can exploit this flexibility to craft sophisticated attacks. Common vectors include Denial-of-Service (DoS) via resource-intensive queries, injection attacks (e.g., SQL, OS commands), Cross-Site Scripting (XSS), schema introspection to find vulnerabilities, and Server-Side Request Forgery (SSRF). Detecting these threats is critical for the security and integrity of GraphQL-based systems.

Existing security solutions like static analysis, rate limiting, and general-purpose Web Application Firewalls (WAFs) are insufficient. They struggle with dynamic attacks, lack a deep understanding of GraphQL's semantics, are difficult to configure correctly, and often do not provide real-time analysis of query payloads. This highlights a clear need for more intelligent, adaptive security mechanisms capable of analyzing dynamic GraphQL queries in real-time.

This paper proposes an AI-driven framework to enhance GraphQL security, integrating machine learning with static analysis to detect vulnerabilities like DoS, injection, and complexity-based attacks. The key contributions are:

\begin{itemize}
    \item A hybrid detection framework tailored for GraphQL's unique structure.
    \item The novel use of Large Language Models (LLMs) to analyze the GraphQL Schema Definition Language (SDL) and dynamically generate context-aware static analysis rules.
    \item The utilization of Sentence Transformers (specifically SBERT for injection attacks and Doc2Vec for XSS) for effective contextual vector embedding of potentially malicious query payloads, enabling pattern-based detection.
    \item The application of Sentence Transformers (SBERT, Doc2Vec) for contextual vector embedding of query payloads to detect injection and XSS attacks.
    \item An evaluation of various machine learning classifiers (CNN, Random Forest, MLP) for accurate payload-based threat detection.
    \item A high-performance, scalable system architecture designed for production environments.
\end{itemize}

The remainder of this paper details the related work, the proposed methodology, experimental results, and a concluding discussion.

\section{Related Work}
GraphQL's flexibility, while efficient for data fetching, introduces unique security vulnerabilities like DoS and injection attacks that traditional API security mechanisms often fail to address. Existing solutions such as static analysis, rate limiting, and general-purpose Web Application Firewalls (WAFs) offer limited protection against sophisticated attacks 1 . Static analysis can miss subtle patterns, rate limiting and depth limits don't inspect content, and WAFs aren't designed for GraphQL's structure. Many current GraphQL security tools focus on one-time server analysis rather than real-time analysis of each incoming query. Examples include GraphQL Cop and Graph00f. There is a clear need for more intelligent and adaptive security mechanisms capable of understanding the context and potential maliciousness of dynamic GraphQL queries in real-time. 

Attack vectors such as SQL injection, OS command injection, and XSS exploits are difficult to detect with static methods as they rely on patterns within queries rather than easily blocked keywords. Machine learning (ML) has emerged as a promising solution for vulnerability detection by enabling the analysis of vast datasets to recognize patterns\cite{b14}.Various ML approaches can be employed effectively. Deep learning (DL), a subfield of ML, is particularly powerful for identifying complex patterns in code and data using architectures like Convolutional Neural Networks (CNNs) and Recurrent Neural Networks (RNNs)\cite{b14}. Transformer-based models, leveraging self-attention mechanisms and handling complex dependencies, show significant potential in cybersecurity for threat detection with high accuracy. They can process sequential and multi-dimensional data effectively\cite{b2}.

The concept of dependency-aware testing is also being explored for GraphQL with tools like GRAPHQLER, which analyze relationships among queries, mutations, and objects to uncover vulnerabilities through context-aware testing\cite{b21}. Despite the potential of ML and DL, challenges remain, including the need for high-quality labeled datasets, managing overfitting, ensuring model interpretability, and adapting to evolving threats\cite{b14}.

This section provides a foundation for understanding the need for advanced techniques, like those combining static analysis with machine learning and natural language processing models, to address the unique security challenges posed by GraphQL APIs in real-time.

\section{Method}
Our AI-driven system uses a hybrid approach for real-time detection of malicious GraphQL queries within an API gateway. It integrates static analysis with machine learning, dynamically configured by the GraphQL schema, and is demonstrated using WSO2 API Manager. This section details the system architecture and its threat detection components.

\subsection{\textbf{Overall System Architecture}}\label{AA}
\begin{figure}[h]
    \centering
    \includegraphics[width=1\linewidth]{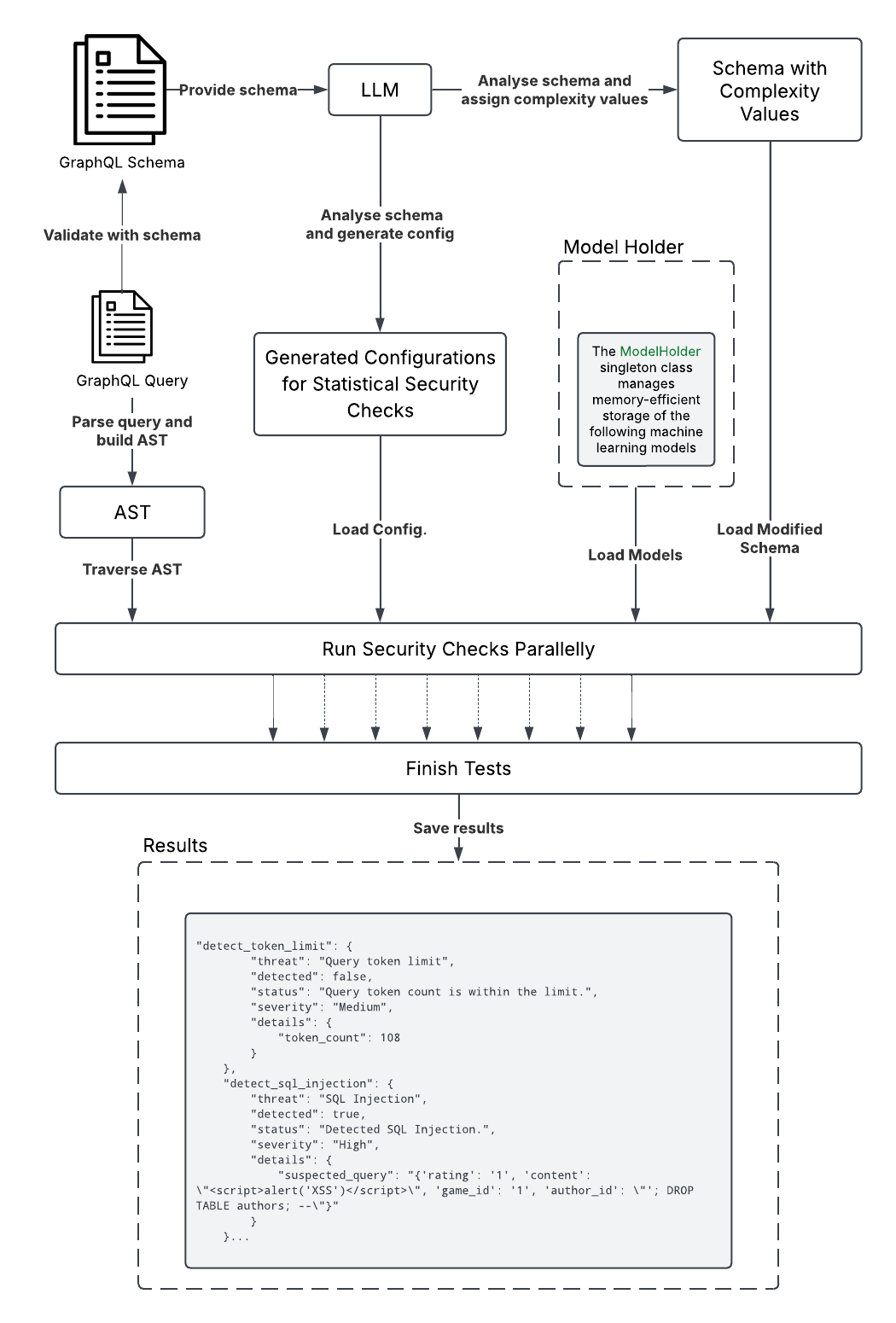}
    \caption{Overall System Architecture}
    \label{fig:overall-archi}
\end{figure}

Our system architecture (Figure \ref{fig:overall-archi}) is built for efficient, parallel processing. Upon receiving a GraphQL query, the system validates it against the schema. Concurrently, an LLM analyzes the schema's SDL based on predefined rules to generate a dynamic configuration file containing thresholds for static checks and complexity values for schema fields.

The query is then parsed into an Abstract Syntax Tree (AST). This AST facilitates the parallel execution of multiple detection modules: static analysis, machine learning inference (for injection and XSS), and SSRF detection. Finally, the results from all modules are aggregated to provide a comprehensive security assessment for the query.

\subsection{\textbf{Smart Dynamic Query Complexity}}
To mitigate Denial-of-Service (DoS) attacks, our system identifies overly complex queries using dynamic schema analysis. We developed two distinct and flexible complexity estimators for this purpose:

\begin{itemize}
    \item \textbf{Simple Estimator}:This estimator calculates query complexity by multiplying a pre-configured fixed value by the query's depth, providing a straightforward, depth-based measure.
    \item \textbf{Directive Estimator}: This sophisticated estimator uses a Large Language Model (LLM) to analyze the GraphQL schema. The LLM intelligently assigns varied complexity values to fields based on their data type and their possible size, while also setting a dynamic complexity threshold according to pre-defined rules. This method provides a more nuanced, context-aware assessment that better reflects actual resource consumption.
\end{itemize}

The user is provided with the option to select their preferred complexity estimator (directive or simple) through a configuration setting, allowing them to tailor the complexity assessment to their specific needs and schema characteristics.

\subsection{\textbf{Thresholding Common DoS Attacks}}
Beyond query complexity, our system mitigates several Denial-of-Service (DoS) vectors by applying dynamic thresholds. A Large Language Model (LLM) generates these thresholds by analyzing the Schema Definition Language (SDL) according to a predefined rule set that dictates how to establish upper bounds for factors like circular dependencies, aliases, and batch sizes. This provides an adaptive defense tailored to the specific API and addresses the following vulnerabilities:

\begin{itemize}
    \item \textbf{Alias Overloading}: Excessive use of field aliases in a query, which increases server processing.
    \item \textbf{Batch Overloading}: A large number of queries sent simultaneously in a single batched request.
    \item \textbf{Deep Circular Queries}: Exploiting cyclical schema relationships to create resource-intensive, deeply nested queries.
    \item \textbf{Directive Overloading}: Overuse of directives (e.g., \texttt{@include}, \texttt{@skip}), which adds significant processing overhead.
    \item \textbf{Excessive Query Depth}:  Crafting queries with an extreme level of nesting that consumes excessive server resources.
    \item \textbf{Query Payload Inflation}: Requesting an overly large volume of data, straining server memory and bandwidth.
\end{itemize}

Upon parsing a query, an Abstract Syntax Tree (AST) is used to traverse its nodes and enforce these LLM-generated, contextual, and meaningful limits.

\subsection{\textbf{Machine Learning for Injection Attack Detection}}
While static analysis is effective for structural vulnerabilities, detecting injection attacks requires a deeper understanding of the content of user inputs. We employ machine learning models for this purpose, leveraging vector embeddings and the handcrafted features.

Injection attacks cannot be identified statically because restricting common keywords for each attack type is impractical. Many benign queries contain these keywords without being harmful.
Therefore, detecting such injection attacks requires analyzing patterns rather than relying solely on keyword-based detection. The following two types of injection attacks are handled in this project effectively.
\begin{itemize}
    \item \textbf{OS Command Injections}: This allows an attacker to execute arbitrary commands on the server where the application is hosted. In a GraphQL context, an OS command injection vulnerability could occur if you have a mutation or query that takes user-supplied input and uses it in a system command
    \item \textbf{SQL injections}: SQL injection (SQLi) attacks are a type of security vulnerability in GraphQL APIs that allow an attacker to execute malicious SQL queries against a backend database. Attackers can exploit this vulnerability by injecting malicious SQL code into a GraphQL query, which is then executed by the backend database.
\end{itemize}

Thus, to identify injection attacks, a machine learning methodology comprised of two primary stages is employed.
\begin{enumerate}
    \item Building contextual vector embeddings for the payloads
    \item Predict vulnerability\\
\end{enumerate}
\subsubsection{\textbf{Phase 01 - Building Contextual Vector Embeddings for the Payloads}}
Various embedding techniques were initially explored for building vector embeddings of the extracted user payloads, including BERT, Microsoft's CodeBERT, Doc2Vec, and FastText from Gensim. Ultimately, the SBERT pre-trained all-MiniLM-L6-v2 model, with 384-dimensional embeddings, demonstrated the best accuracy on the final validation set.

\begin{figure}[h]
    \centering
    \includegraphics[width=0.8\linewidth]{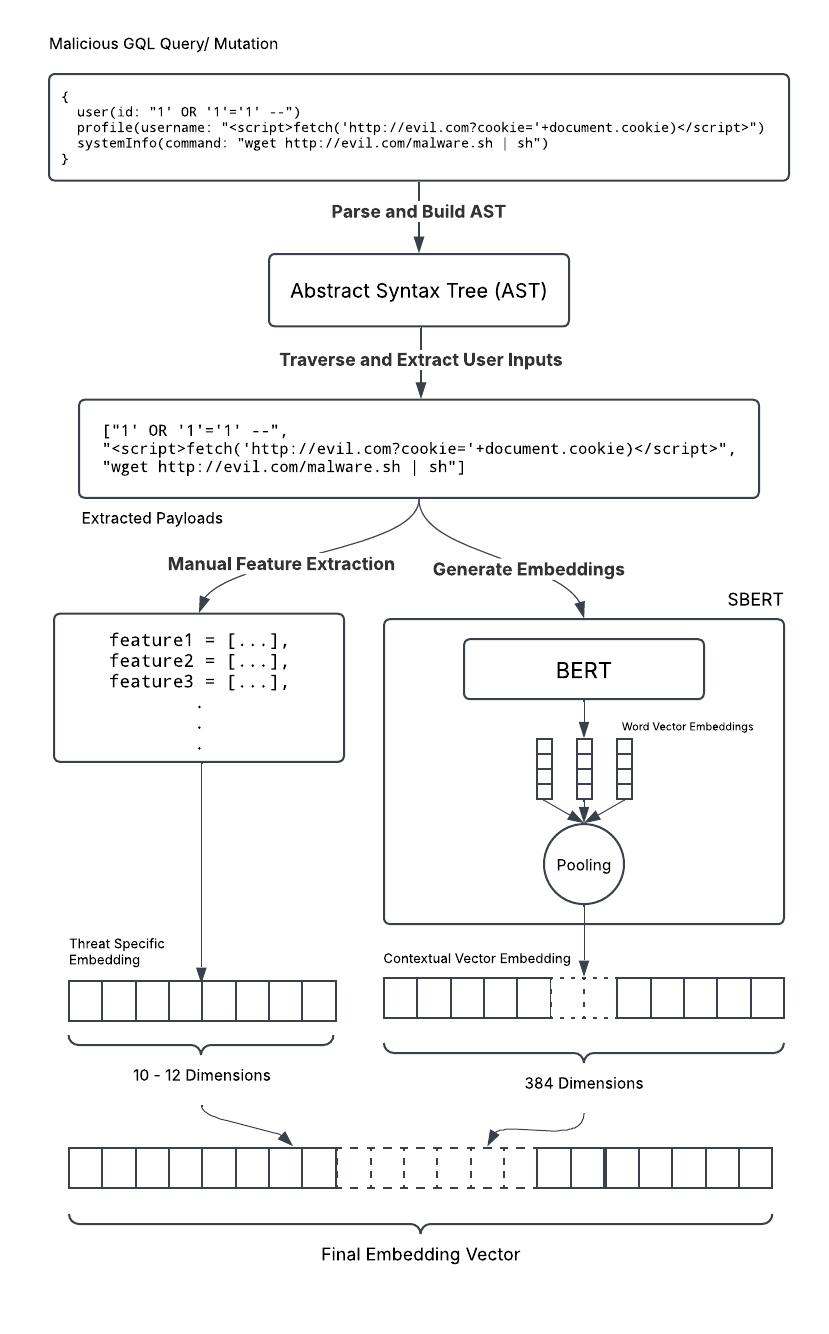}
    \caption{Building Contextual Vector Embeddings for the Payloads with SBERT}
    \label{fig:sbert}
\end{figure}

As shown in Figure \ref{fig:sbert}, our model for injection attacks first generates a base embedding vector from the input. This vector is then appended with specific handcrafted features to create two distinct vectors, one tailored for SQL Injection (SQLi) and one for OS command injection. Each vector is subsequently processed by a dedicated 1D CNN. Although the CNNs share the same architecture, each is trained on its corresponding attack-specific dataset, resulting in two specialized models for injection detection.\\

\textbf{Handcrafted Features for OS Injection Detection}
\begin{itemize}
    \item \textbf{OS Commands Count} – Counts occurrences of common OS commands like \texttt{ls}, \texttt{pwd}, \texttt{chmod}, and \texttt{whoami}.
    \item \textbf{OS Operators Count} – Detects command chaining operators (\texttt{|}, \texttt{\&\&}, \texttt{;}, \texttt{>}, \texttt{>>}) that facilitate injection.
    \item \textbf{OS Special Characters Count} – Identifies special shell characters (\texttt{\$}, \texttt{\&}, \texttt{;}, \texttt{\{\}}, \texttt{()}) often used in exploits.
    \item \textbf{OS Payload Patterns} – Matches known OS command injection payloads, including remote shells and command execution patterns.
    \item \textbf{Pipe Operators Count} – Tracks the use of \texttt{||}, \texttt{\&\&}, and \texttt{|}, which can be used to append or manipulate commands.
    \item \textbf{Variable Execution Count} – Detects command substitution using \texttt{\$()} and backticks (\texttt{`...`}), which execute embedded commands.
    \item \textbf{Remote Execution Keywords} – Identifies commands used for remote file retrieval and execution (\texttt{wget}, \texttt{curl}, \texttt{scp}).
    \item \textbf{System Information Extraction} – Flags queries attempting to extract system details using \texttt{uname}, \texttt{whoami}, and \texttt{env}.
    \item \textbf{Privilege Escalation Attempts} – Detects commands (\texttt{sudo}, \texttt{su}, \texttt{chmod}, \texttt{chown}) that attempt to escalate privileges.\\
\end{itemize}

\textbf{Handcrafted Features for SQL Injection Detection}

\begin{itemize}
    \item \textbf{SQL Keywords Count} – Counts occurrences of critical SQL commands (\texttt{INSERT}, \texttt{UPDATE}, \texttt{DELETE}, \texttt{DROP}, \texttt{ALTER}).
    \item \textbf{SQL Operators Count} – Identifies SQL-specific operators (\texttt{--}, \texttt{/*}, \texttt{*/}, \texttt{@@}, \texttt{CAST}, \texttt{CONVERT}).
    \item \textbf{SQL Special Characters Count} – Detects special characters (\texttt{'}, \texttt{"}, \texttt{--}, \texttt{;}, \texttt{/*}) often used in SQL injection payloads.
    \item \textbf{Boolean Conditions Count} – Flags logical SQL conditions (\texttt{XOR}, \texttt{NOT}) that can be used for bypassing authentication.
    \item \textbf{Query Length} – Measures the total length of user input, as longer inputs may indicate payloads.
    \item \textbf{Union-Select Usage Count} – Tracks usage of \texttt{UNION} and \texttt{SELECT}, commonly seen in injection attempts.
    \item \textbf{SQL Payload Patterns Count} – Matches known SQL injection payloads like \texttt{' OR '1'='1' --} and \texttt{DROP TABLE}.
    \item \textbf{Encoded Injection Count} – Detects percent-encoded (\texttt{\%27}, \texttt{\%3D}, \texttt{0x27}) SQL payloads used for obfuscation.
    \item \textbf{Database-Specific Keywords Count} – Flags DB-specific functions (\texttt{information\_schema}, \texttt{xp\_cmdshell}, \texttt{database()}).
    \item \textbf{Time-Based Attack Keywords Count} – Identifies delay-based injection techniques (\texttt{SLEEP}, \texttt{BENCHMARK}, \texttt{WAITFOR DELAY}).
    \item \textbf{Nested Select Count} – Detects nested \texttt{SELECT} queries, which can be used to manipulate database logic.\\
\end{itemize}

\subsubsection{\textbf{Phase 02 - Predict Vulnerability}}
\begin{figure*}
    \centering
    \includegraphics[width=1\linewidth]{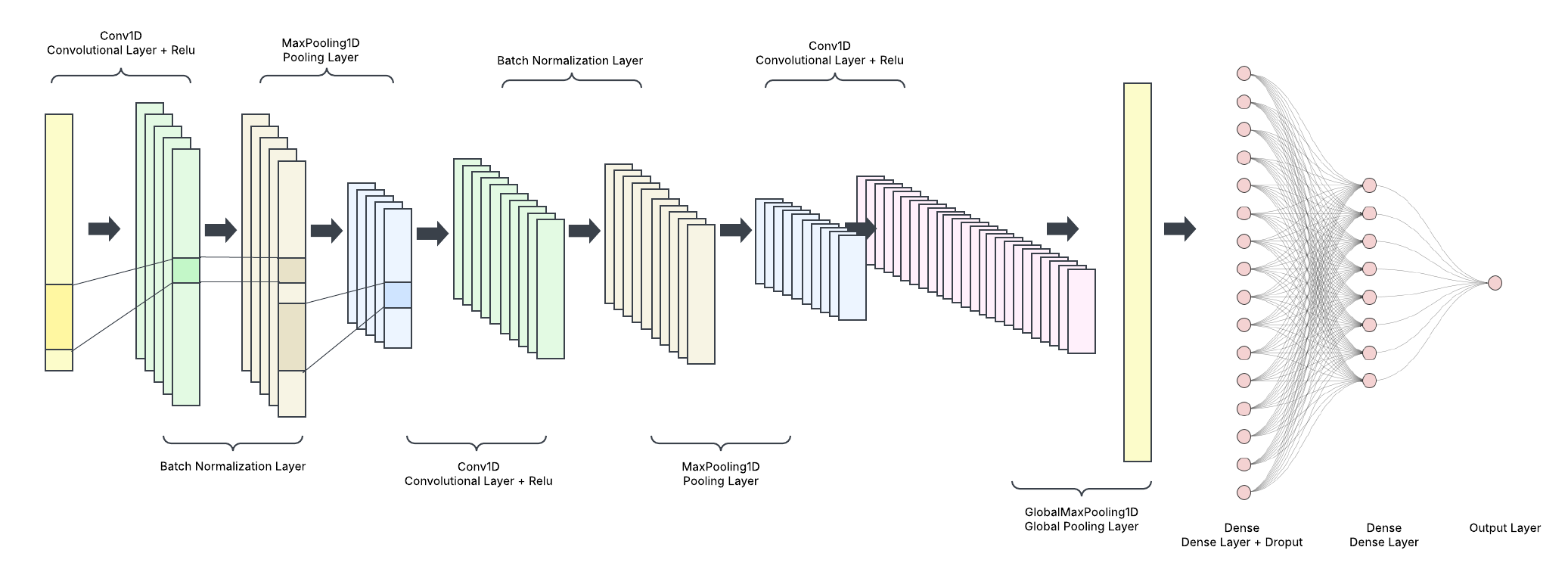}
    \caption{CNN Architecture}
    \label{fig:cnn}
\end{figure*}
The concatenated feature vector is classified by a 1D CNN, with identical models for SQLi and OS Command detection. As shown in Figure \ref{fig:cnn}, it has three 1D convolutional layers (128, 256, 512 filters; kernel 3; ReLU), each with Batch Normalization and MaxPooling1D (2). A GlobalMaxPooling1D layer precedes a 256-neuron dense layer with 0.5 dropout, and a final sigmoid neuron outputs malicious probability. The model uses Adam (0.001) with binary cross-entropy, trained up to 20 epochs (batch 32) with early stopping (5 patience) and best-weight restoration.

\subsection{\textbf{Machine Learning for XSS Exploit Detection}}
Like injection attacks, XSS (Cross-Site Scripting) exploits cannot be identified simply by blocking specific keywords. In this case too, a two-step machine learning approach has been employed to identify such vulnerabilities.
\begin{enumerate}
    \item Building contextual vector embeddings for the payloads
    \item Predict vulnerability\\
\end{enumerate}

\subsubsection{\textbf{Phase 01 - Building Contextual Vector Embeddings for the Payloads}}
XSS detection is easier than injection attacks due to common exploit patterns. Injection attacks are more complex and diverse, so high-dimensional embeddings are unnecessary for XSS and would add extra overhead.

\begin{figure}[h]
    \centering
    \includegraphics[width=0.8\linewidth]{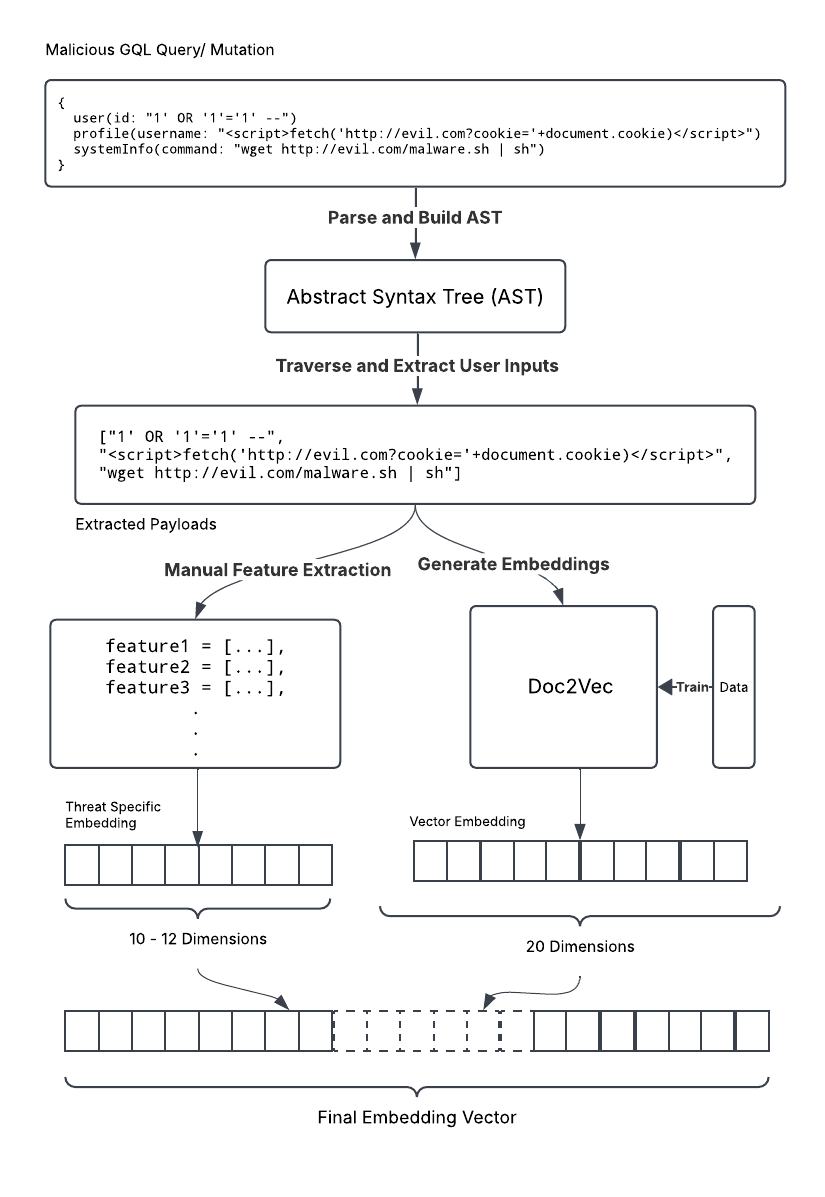}
    \caption{Building contextual vector embeddings for the payloads with Doc2Vec}
    \label{fig:doc2vec}
\end{figure}

To optimize performance, as shown in Figure \ref{fig:doc2vec}, the Doc2Vec embedding model from Gensim is used, generating an embedding vector of size 20. The Doc2Vec model employed in this study is a custom-trained version, developed specifically using a dataset of XSS and benign scripts, rather than a pre-trained model. This vector is then concatenated with additional handcrafted features extracted from the query. The combined feature vector is subsequently fed into the models to determine whether the query is malicious or benign.

\textbf{Handcrafted Features for XSS Exploit Detection}
\begin{itemize}
    \item \textbf{HTML Tag Count} – Counts occurrences of potentially dangerous HTML tags (\texttt{<script>}, \texttt{<img>}, \texttt{<iframe>}, \texttt{<input>}, etc.) often used in XSS payloads.
    \item \textbf{JavaScript Method Count} – Identifies use of sensitive JavaScript functions (\texttt{eval()}, \texttt{alert()}, \texttt{document.write()}, etc.) commonly exploited in XSS attacks.
    \item \textbf{.js File Reference Count} – Tracks references to external JavaScript files (e.g., \texttt{.js}) which may contain malicious code.
    \item \textbf{Keyword "javascript" Count} – Measures the frequency of the keyword \texttt{javascript}, often used in URI-based or inline scripts.
    \item \textbf{Payload Length} – Represents the total length of the payload; longer inputs may include embedded scripts or obfuscation.
    \item \textbf{Obfuscated Script Variants Count} – Flags encoded or obfuscated versions of \texttt{<script>} (e.g., \texttt{\%3Cscript}, \texttt{\&lt;script}) used to bypass filters.
    \item \textbf{Special Characters Count} – Counts occurrences of characters like \texttt{<}, \texttt{>}, \texttt{"}, \texttt{'} and encoded forms (\texttt{\%3C}, \texttt{\%3E}) that are critical in HTML/JS injection.
    \item \textbf{External Resource Count} – Detects usage of URLs (\texttt{http}) that may load remote scripts or redirect users to malicious domains.\\
\end{itemize}

\subsubsection{\textbf{Phase 02 - Predict Vulnerability}}
To enhance the efficiency of malicious XSS detection, an ensamble method of a Random Forest classifier and a Multilayer Perceptron (MLP) were employed. This approach was chosen due to the reduced dimensionality of the handcrafted and embedded feature vectors, which makes traditional machine learning models particularly suitable. Unlike deep neural networks that often require large amounts of data and computational resources, these models offer faster training and inference with minimal performance trade-offs. Moreover, this design helps significantly reduce computational overhead while maintaining high detection accuracy. The ensemble of Random Forest and MLP achieves an effective balance between precision and efficiency.

\subsection{\textbf{Mitigating Server-side Request Forgery (SSRF) Attempts}}
To mitigate Server-Side Request Forgery (SSRF) attacks, where adversaries manipulate server-side requests, our system performs several parallel security checks. Upon receiving a query, its Abstract Syntax Tree (AST) is inspected for URLs. Any URL found is then evaluated against the following attack vectors:

\subsubsection{\textbf{Local IP Attack Detection}}

This check blocks requests to local addresses (\texttt{localhost}, \texttt{127.0.0.1}), private IP ranges, and their obfuscated variants (e.g., encoded IPs, DNS redirects) to prevent access to internal-only network resources.

\subsubsection{\textbf{Cloud Metadata Attack Prevention}}
This prevents access to well-known metadata service endpoints for major cloud providers (AWS, GCP, Azure) by blocking requests to their specific IPs and hostnames, thus protecting sensitive cloud credentials.

\subsubsection{\textbf{Parameter Based SSRF Prevention}}
We scan for query parameters commonly abused in SSRF (e.g., \texttt{url}, \texttt{redirect}). If these parameters contain a URL, it is subjected to our other SSRF checks, preventing attackers from hiding malicious URLs in trusted parameters.

\subsubsection{\textbf{Encoded Payload Attack Mitigation}}
To counter evasion techniques, the system detects malicious URLs that are obfuscated using encoding schemes like URL encoding, Base64, or Unicode by checking them against a list of known encoded attack patterns.

\subsection{\textbf{Implementation and Optimization for Production}}
The detection system is implemented as a modular, high-performance service designed for real-time analysis within a production environment. This section details the key implementation strategies and optimizations employed to ensure the system's efficiency, scalability, and seamless integration.

\subsubsection{\textbf{Asynchronous Architecture}}
The core service is built on FastAPI, an asynchronous framework, to efficiently handle a high volume of concurrent requests without blocking. This is crucial for processing real-time API traffic.
\subsubsection{\textbf{Centralized Model Loader}}
To minimize memory and loading overhead, a singleton class manages all machine learning models. The models are loaded only once upon system initialization, making them readily available for inference across all subsequent requests.
\subsubsection{\textbf{ONNX Runtime Optimization}}
\begin{figure}[h]
    \centering
    \includegraphics[width=1\linewidth]{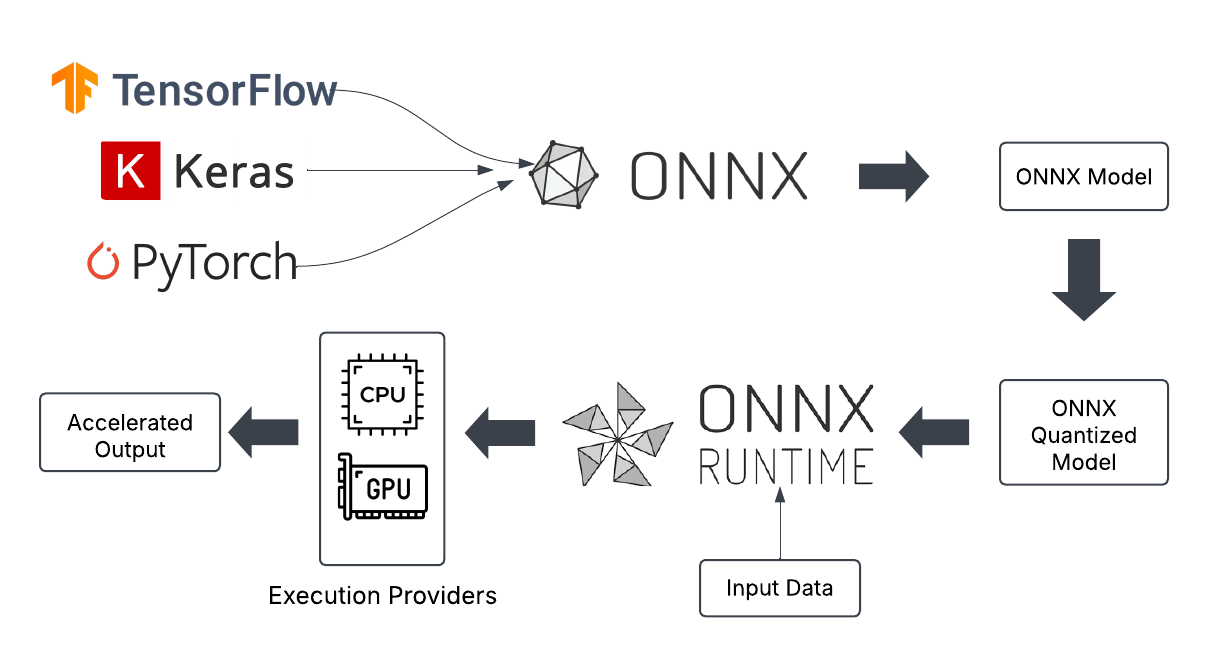}
    \caption{ONNX Runtime Optimization}
    \label{fig:onnx}
\end{figure}

For maximum inference speed, all trained models are converted to the Open Neural Network Exchange (ONNX) format and quantized to INT8 precision.  We use the ONNX Runtime, configured with aggressive graph optimizations, to execute these models (Figure \ref{fig:onnx}). This significantly accelerates predictions by reducing memory footprint and leveraging hardware-level optimizations.

\subsubsection{\textbf{Optimized Parsing and Preprocessing}}
GraphQL query parsing and the expansion of fragments are performed only once per query in an initial preprocessing stage. This prevents redundant computation by ensuring the Abstract Syntax Tree (AST) is built efficiently before being passed to the various detection modules.

\subsubsection{\textbf{Parallel and Concurrent Vulnerability Detection}}
The system uses a hybrid execution model to maximize throughput. A dedicated thread pool handles CPU-intensive tasks like ML model inference, while another manages I/O-bound operations. The main \texttt{asyncio} event loop coordinates these tasks, ensuring the application remains responsive while efficiently processing computationally intensive or blocking operations in separate threads.

\subsubsection{\textbf{Centralized Logging}}
A custom logging system provides clear visibility into all key operations while suppressing verbose output from external libraries. To minimize I/O overhead and reduce latency, logs are written to a file in batches.

\subsubsection{\textbf{Production Deployment Architecture}}
For production, the FastAPI application runs using Gunicorn as a process manager with Uvicorn workers.  This architecture leverages Gunicorn for robust concurrency and process supervision, while Uvicorn provides a high-performance Asynchronous Server Gateway Interface (ASGI) for the application, ensuring scalability under substantial user load.

\subsection{\textbf{Data Collection and Preparation}}
This section outlines the methodologies employed for gathering and preparing the datasets utilized in the training and evaluation of the machine learning models. Separate datasets were compiled for each targeted attack vector: SQL Injection, OS Command Injection, and Cross-Site Scripting (XSS).

\subsubsection{\textbf{Data Acquisition}}
The data acquisition phase involved collecting relevant payloads for each attack type from established sources.

For SQL Injection detection, a dataset was obtained from a publicly available repository on Kaggle\cite{b30}. The OS Command Injection dataset was acquired from the resource specified in\cite{b27}. In the case of XSS exploits, diverse payloads were gathered from the sources indicated by\cite{b28},\cite{b29},\cite{b31}, and \cite{b32}.
The approximate volumes of the datasets after cleaning, are as mentioned in Table I.

\begin{table}[h]
    \renewcommand{\arraystretch}{1.3}
    \centering
    \caption{Approximate Volumes of Datasets}
    \label{tab:my_label}
    \begin{tabular}{|c|c|c|}
        \hline
        \textbf{Dataset} & \textbf{Malicious} & \textbf{Benign} \\
        \hline
         SQL Injection & 77k  & 75k \\
         OS Command Injection & 7.5k & 7.5k \\
         XSS & 38k & 44k \\
         \hline
    \end{tabular}
\end{table}

\subsubsection{\textbf{Data Pre-processing}}
Datasets were preprocessed to ensure consistency, with each instance containing a \texttt{payload} and binary \texttt{label} (1 for malicious, 0 benign). Hundreds of mislabeled samples were removed after manual review. For the limited OS Command Injection data, LLM-based augmentation generated synthetic payloads combining OS commands and natural language to improve robustness.

Negative sampling incorporated samples from other attack categories labeled benign within each dataset to reduce false positives caused by similar code-like patterns across categories, enhancing model specificity for SQL Injection, OS Command Injection, and XSS detection.

\section{Results}
This section presents the evaluation results of our AI-driven GraphQL security detection system. We first detail the performance of the machine learning models developed for detecting injection attacks and XSS exploits. Subsequently, we present the results of load testing and system profiling conducted to assess the system's performance and scalability in a simulated production environment.

\subsection{\textbf{Machine Learning Model Evaluation}}

The ML models for SQL Injection, OS Command Injection, and XSS detection were evaluated on a held-out test set (Section III.H) using Accuracy, Precision, Recall, and F1-score. Confusion matrices detailed true/false positives and negatives. Results are summarized in Tables II and III with corresponding confusion matrices.


\begin{table}[h]
    \renewcommand{\arraystretch}{1.3}
    \centering
    \caption{Performance Metrics for Injection Attack Detection Models}
    \label{tab:placeholder}
    \begin{tabular}{|c|c|c|}
        \hline
        \textbf{Model} & \textbf{CNN for SQLi} & \textbf{CNN for OS Command} \\ \hline
        Accuracy       & 0.9678 & 0.9767 \\ \hline
        Precision      & 0.9940 & 0.9950 \\ \hline
        Recall         & 0.9403 & 0.9659 \\ \hline
        F1-score       & 0.9664 & 0.9802 \\ \hline
        Confusion Matrix &
        \includegraphics[width=2.5cm]{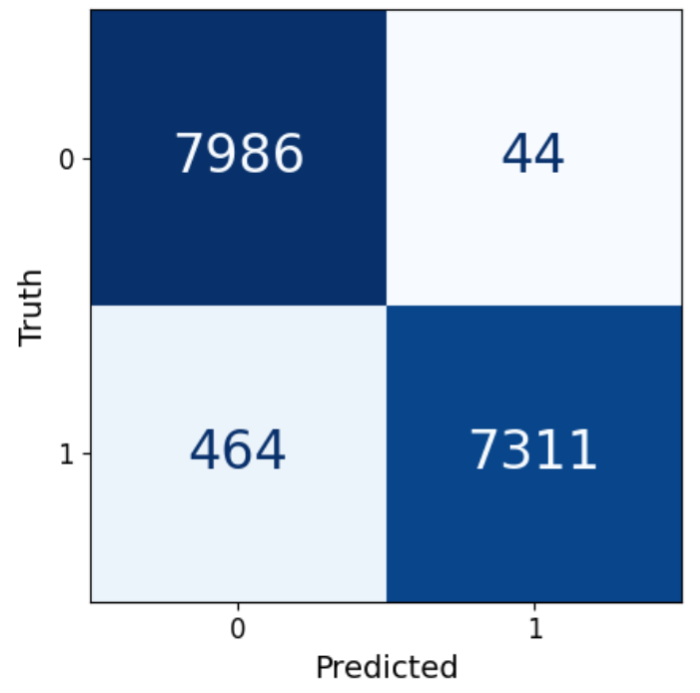} &
        \includegraphics[width=2.5cm]{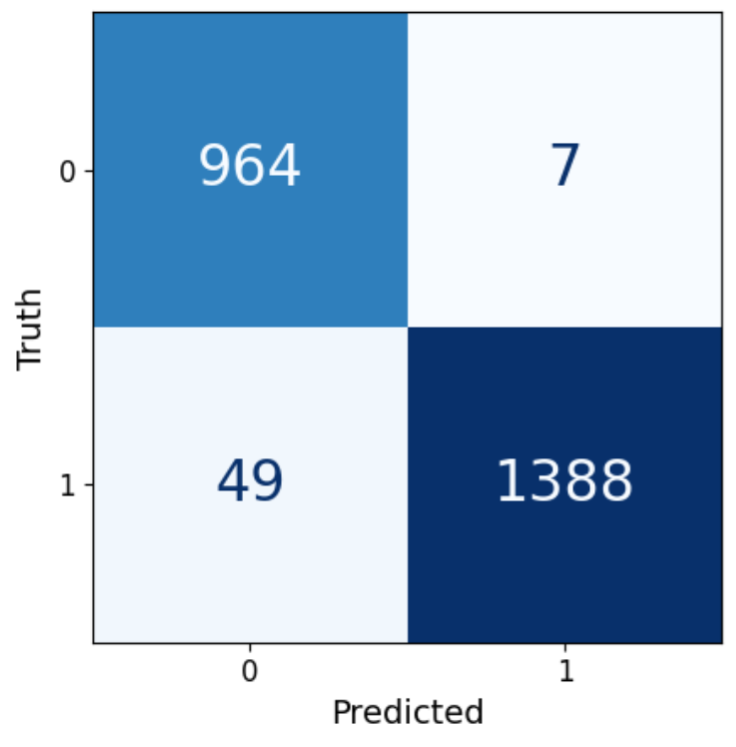} \\ \hline
    \end{tabular}
\end{table}

Overall, both SQL Injection and OS Command Injection models achieved high accuracy and robust detection, with SBERT embeddings and handcrafted features, processed by dedicated CNNs, effectively capturing the complex patterns of malicious payloads.

\begin{table}[h]
    \renewcommand{\arraystretch}{1.3}
    \centering
    \caption{Performance Metrics for XSS Exploit Detection Models}
    \label{tab:placeholder}
    \begin{tabular}{|c|c|c|}
        \hline
        \textbf{Model} & \textbf{Random Forest Classifier} & \textbf{MLP Classifier} \\ \hline
        Accuracy       & 0.9938 & 0.9948 \\ \hline
        Precision      & 0.9988 & 0.9961 \\ \hline
        Recall         & 0.9879 & 0.9926 \\ \hline
        F1-score       & 0.9933 & 0.9943 \\ \hline
        Confusion Matrix &
        \includegraphics[width=2.5cm]{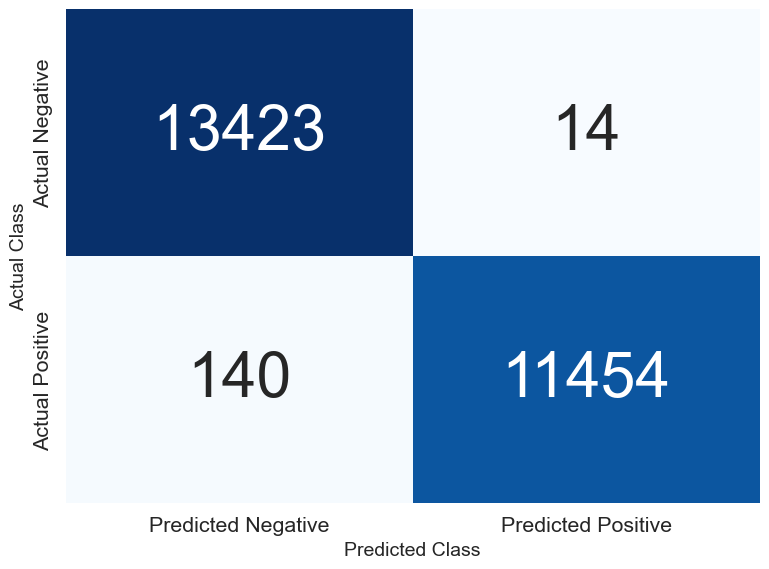} &
        \includegraphics[width=2.5cm]{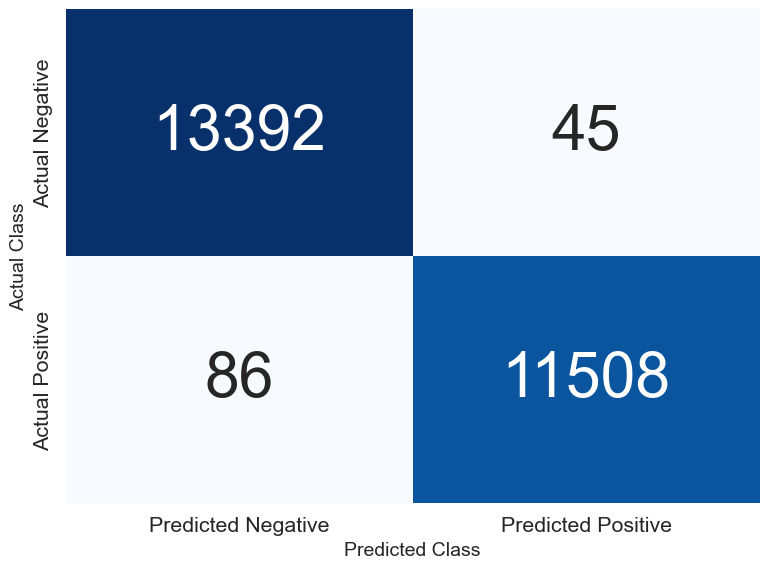} \\ \hline
    \end{tabular}
\end{table}

Both the Random Forest and MLP models demonstrated high accuracy in identifying XSS exploits, indicating that the Doc2Vec embeddings combined with handcrafted features are effective for this type of threat.

\subsection{\textbf{System Performance and Scalability}}
To assess the system's performance and scalability under load, we conducted load testing in a simulated production environment.

\subsubsection{\textbf{Load Testing Methodology and Environment}}
Load testing was performed within an Azure Virtual Machine environment with the following specifications: 
\begin{table}[h]
    \centering
    \begin{tabular}{ll}
        \hline
        \textbf{Spec}               & \textbf{Details}         \\
        \hline
        Cloud Provider                   & Microsoft Azure          \\
        VM Series                        & F-Series                 \\
        vCPUs                            & 4                        \\
        Memory                           & 8 GB                     \\
        Disk Type and Size               & 64 GB HDD                \\
        Operating System                 & 22.04.1-Ubuntu           \\
        Max Network Bandwidth (Mb/s)     & 10000                    \\
        Accelerators                     & None                     \\
        \hline
    \end{tabular}
\end{table}

Locust was used to simulate concurrent load by sending POST requests with mixed benign and malicious GraphQL queries to the API endpoint. The test ramped from 0 to 500 users at 10 users/sec over 2 minutes. Load testing was performed in two phases to assess the performance of different components.
\subsubsection{\textbf{Benchmarking Static Security Checks}}
\begin{figure}[h]
    \centering
    \includegraphics[width=0.77\linewidth]{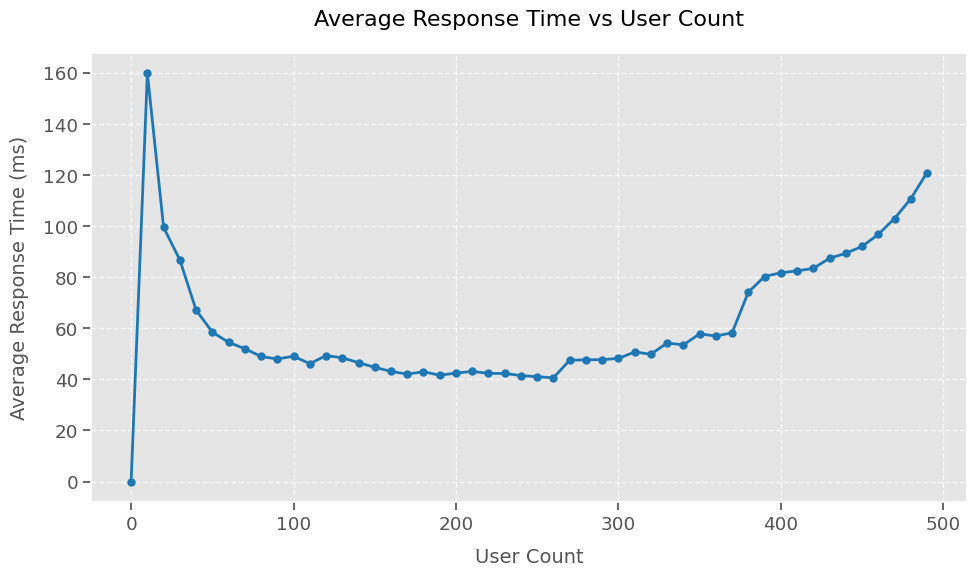}
    \caption{Average Response Time vs User Count in Static Checks}
    \label{fig:st1}
\end{figure}

\begin{figure}[h]
    \centering
    \includegraphics[width=0.77\linewidth]{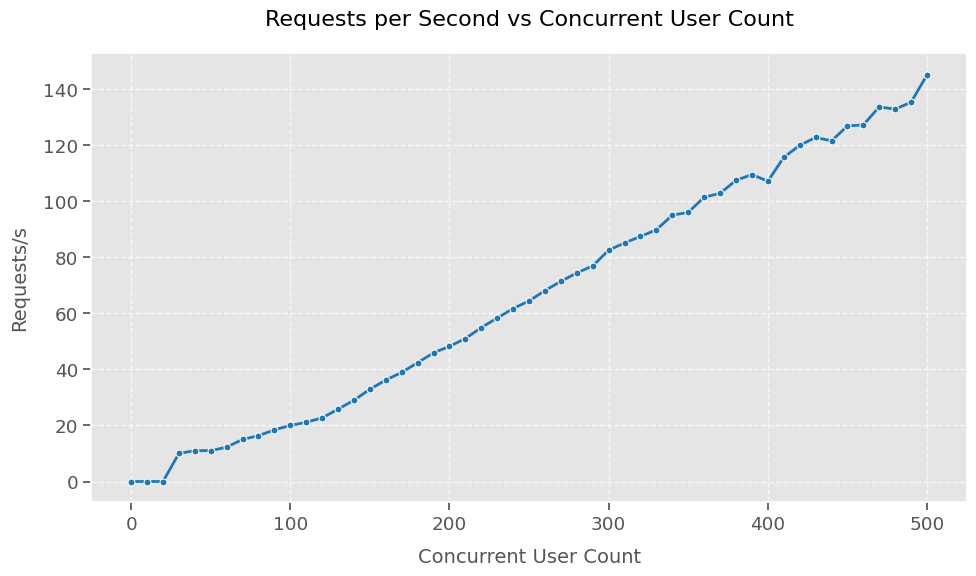}
    \caption{Requests per Second vs Concurrent User Count in Static Checks}
    \label{fig:st2}
\end{figure}

\begin{figure}[h]
    \centering
    \includegraphics[width=0.8\linewidth]{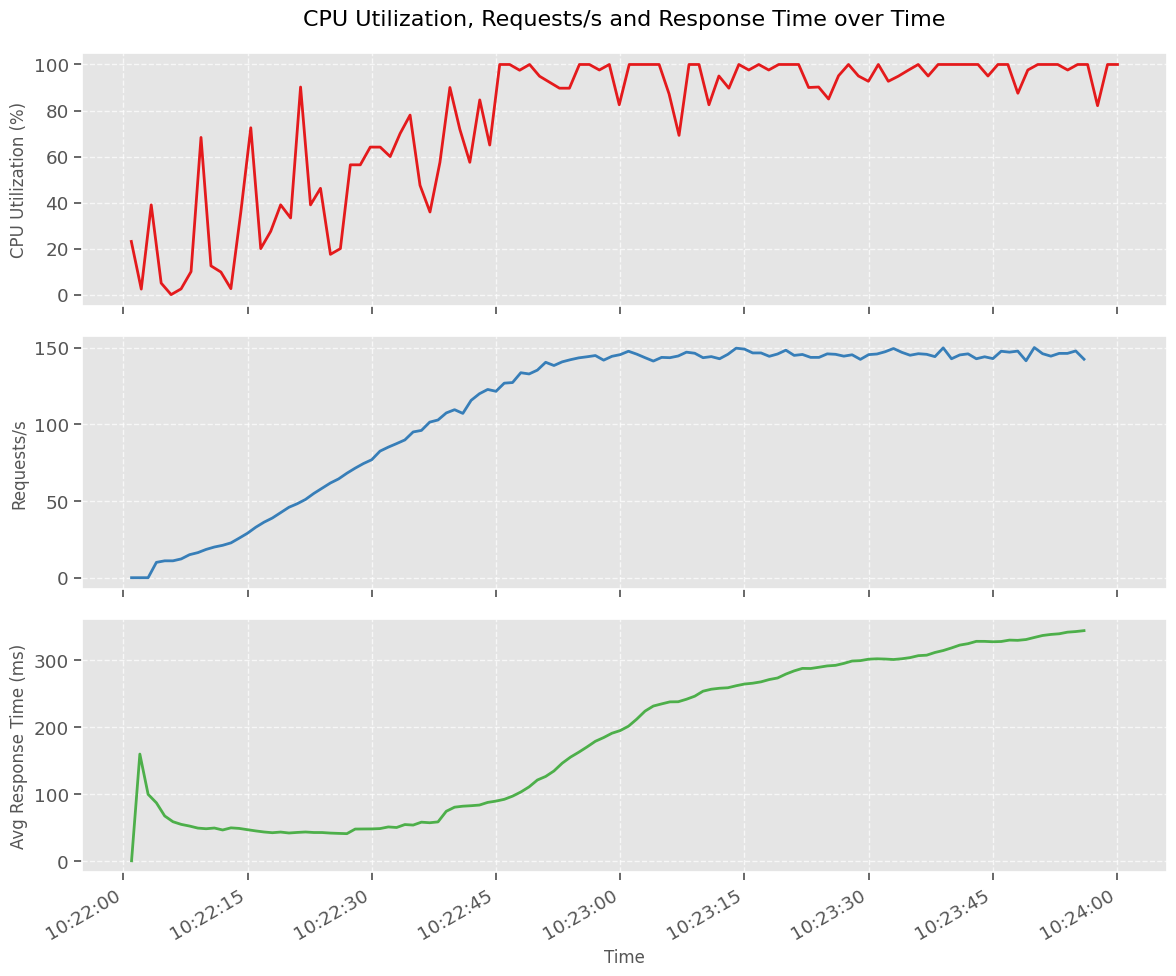}
    \caption{CPU Utilization, Requests/s and Response Time over Time in Static Checks}
    \label{fig:st3}
\end{figure}

With only static checks (Alias Overloading, Batch Overloading, Deep Circular Queries, Directive Overloading, Introspection Queries, Excessive Query Complexity/Depth, Query Payload Inflation, SSRF) enabled, the system sustained high concurrency with low average response times (Figures \ref{fig:st1}, \ref{fig:st2}, \ref{fig:st3}), demonstrating the computational efficiency of static analysis and thresholding mechanisms.

\subsubsection{\textbf{Benchmarking All Security Checks}}
\begin{figure}[h]
    \centering
    \includegraphics[width=0.8\linewidth]{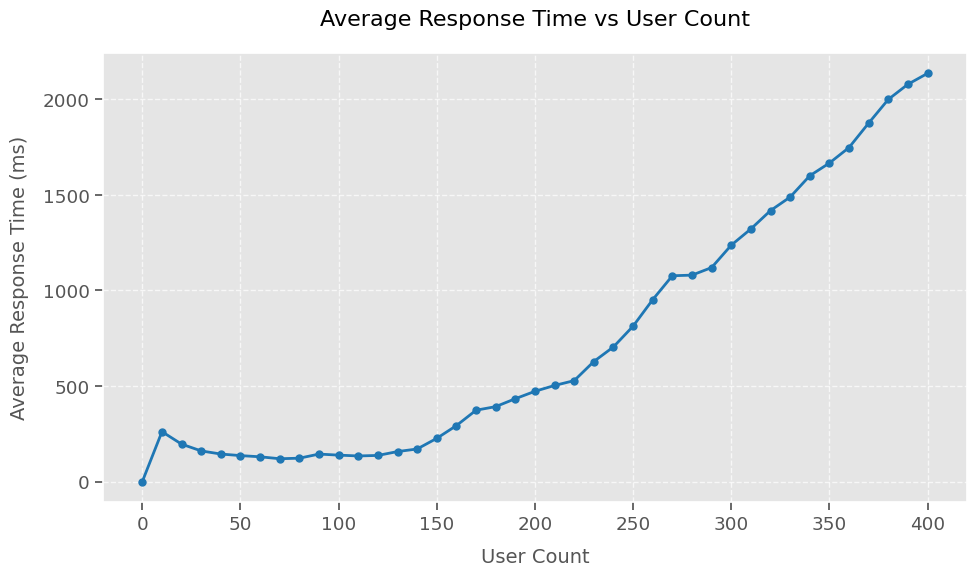}
    \caption{Average Response Time vs User Count for All Checks}
    \label{fig:all1}
\end{figure}

\begin{figure}[h]
    \centering
    \includegraphics[width=0.8\linewidth]{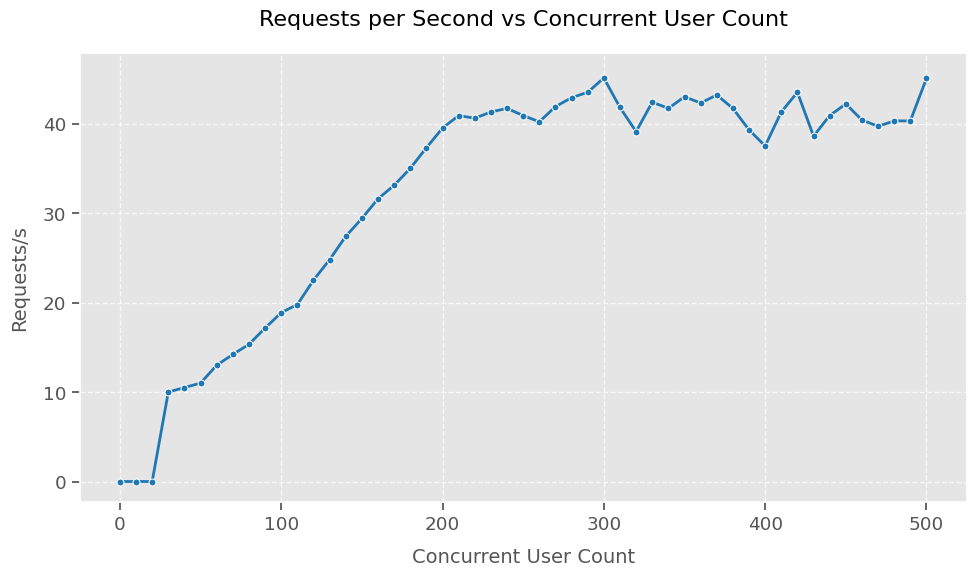}
    \caption{Requests per Second vs Concurrent User Count for All Checks}
    \label{fig:all2}
\end{figure}

\begin{figure}[h]
    \centering
    \includegraphics[width=0.8\linewidth]{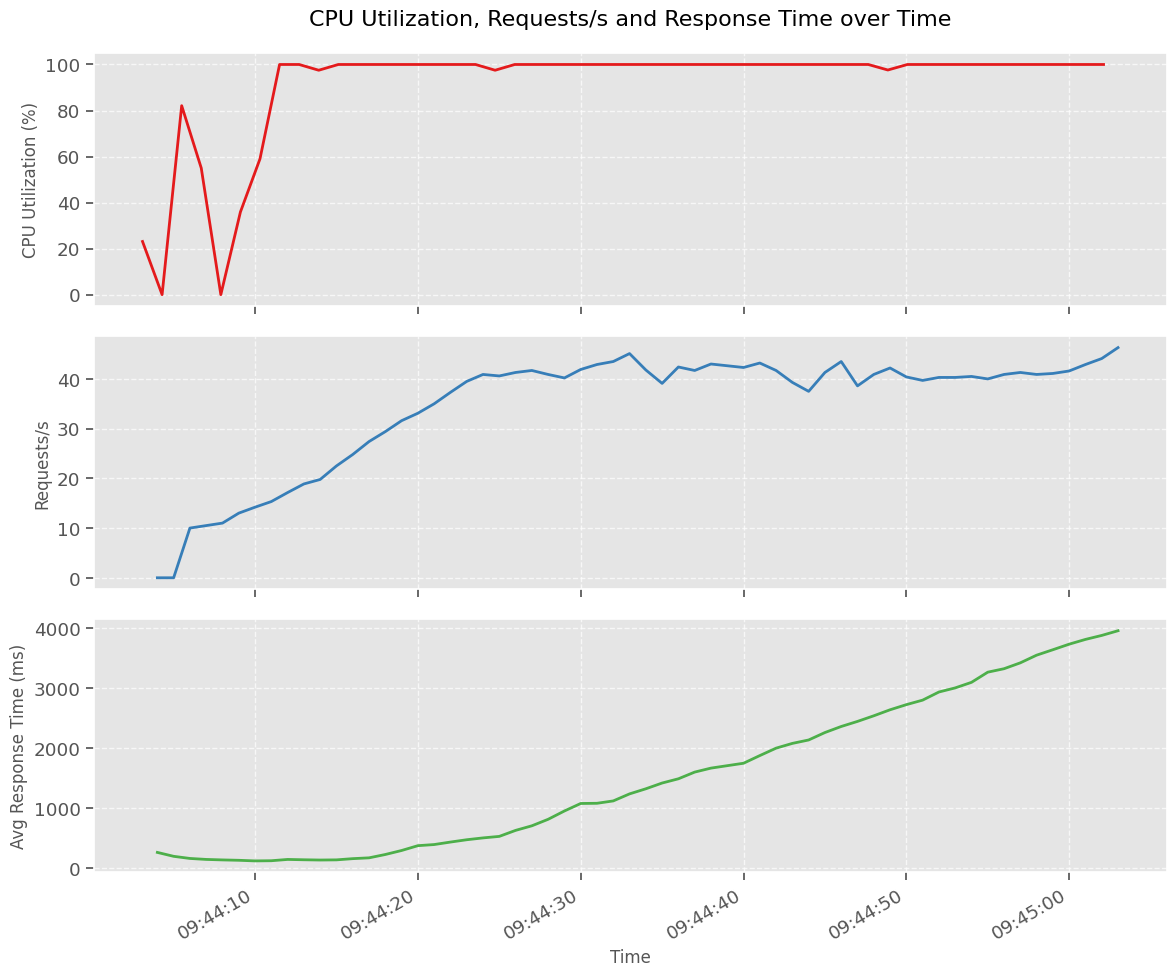}
    \caption{CPU Utilization, Requests/s and Response Time over Time for All Checks}
    \label{fig:all3}
\end{figure}

With all checks enabled, including ML-based SQL Injection, OS Command Injection, and XSS detection, performance metrics (Figures \ref{fig:all1}, \ref{fig:all2}, \ref{fig:all3}) showed notable overhead. Average response time rose sharply with higher concurrency, mainly due to the computational cost of ML inference. All tests were conducted in a non-GPU environment.

\subsubsection{\textbf{System Profiling}}
\begin{figure}[h]
    \centering
    \includegraphics[width=1\linewidth]{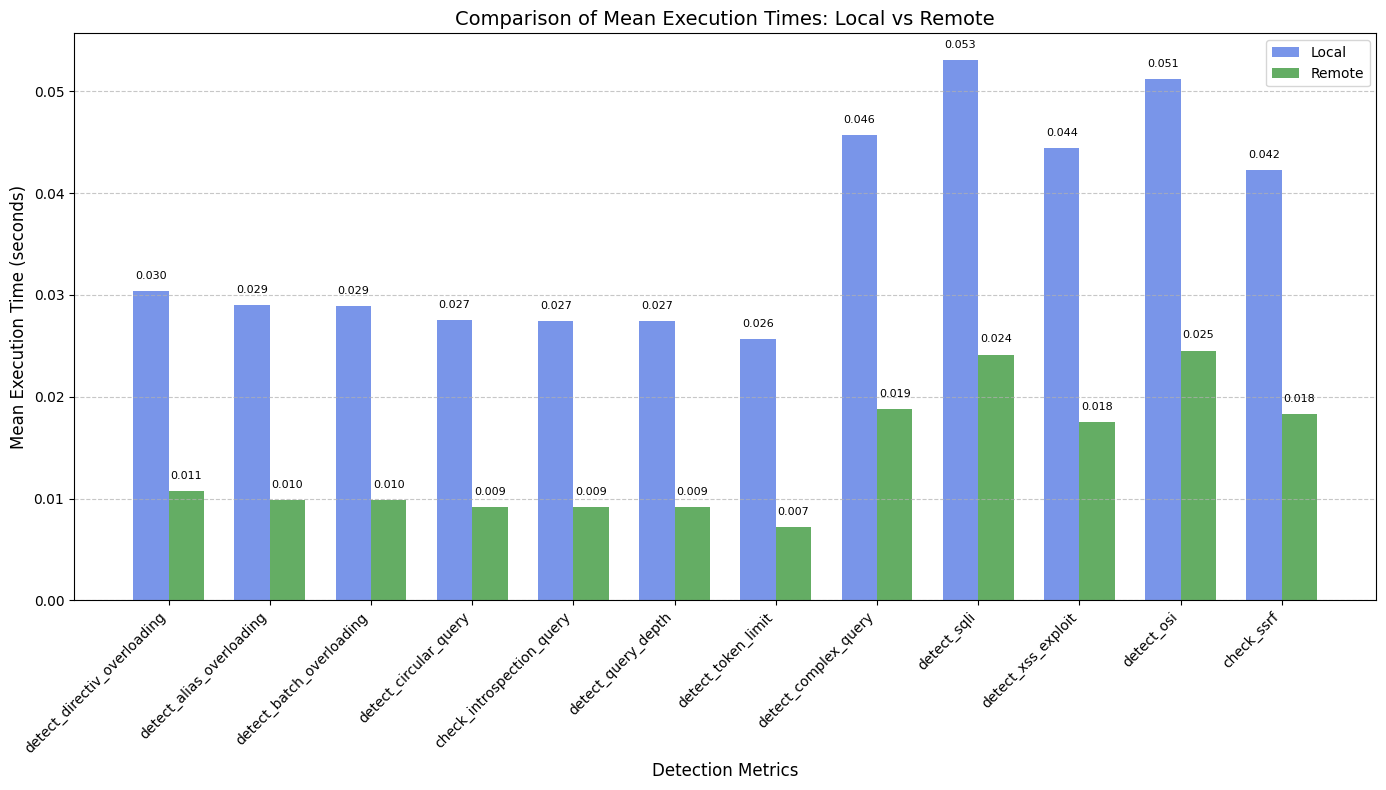}
    \caption{Security Checks Execution Metrics}
    \label{fig:exec-stats}
\end{figure}

System profiling in local and remote environments measured execution times for each security check. As shown in Figure \ref{fig:exec-stats}, ML inference tasks (\texttt{detect\_sqli}, \texttt{detect\_xss\_exploit}, \texttt{detect\_osi}) were significantly slower than static checks, while SSRF (\texttt{check\_ssrf}) also took longer due to AST traversal. The main overhead with all checks enabled stems from CPU-bound ML inference, especially in the test setup lacking GPUs and with limited CPU cores. Running multiple AI-based checks in parallel increased response times under high concurrency. Although ONNX Runtime optimization and parallel execution (Section III.G) reduced this, profiling confirms ML inference remains the primary bottleneck in CPU-only environments.

\section{Discussion}
This section evaluates the AI-driven GraphQL security detection system. The hybrid approach—combining SBERT/Doc2Vec embeddings with handcrafted features—achieves high accuracy (Tables 1–2) in detecting injection and XSS attacks, even when obfuscated. SBERT captures semantic meaning, while handcrafted features detect known patterns.

Load testing shows a performance trade-off: static analysis remains efficient under high concurrency (Figures 8–10), but ML inference introduces significant overhead (Figures 11–13), with profiling confirming CPU-only environments as a bottleneck (Figure 14). Despite ONNX and parallel execution optimizations, hardware acceleration (GPU/TPU) is needed for production scalability.

The LLM’s dynamic thresholding reduces manual configuration by adapting static checks to various schemas, though further quantitative evaluation is required to confirm its effectiveness across diverse scenarios.

\section{Conclusion}
This paper introduced a novel, AI-driven system for real-time GraphQL security. Our hybrid approach, which combines static analysis with machine learning, leverages LLMs for dynamic configuration and Sentence Transformers with CNNs for effective threat detection.

The evaluation demonstrated high accuracy in detecting SQLi, OS Command, and XSS exploits. However, load testing revealed that real-time ML inference introduces computational overhead, particularly in CPU-only environments, underscoring the need for hardware acceleration.

Our work provides a robust and adaptable solution for GraphQL API security. The dynamic configuration via LLMs allows for schema-specific policies, while the combined static and ML checks provide comprehensive threat coverage.

Future work will focus on expanding the dataset, optimizing models, evaluating performance on accelerated hardware, incorporating XAI techniques, and exploring advanced feedback mechanisms for continuous improvement.

\end{document}